\documentclass[preprint,showpacs,preprintnumbers,amsmath,amssymb]{revtex4}
\newcommand{\newwidth}{0.675\textwidth}
\newcommand{\newheight}{0.45\textwidth}
\newcommand{\newwidthprime}{0.225\textwidth}
\newcommand{\newheightprime}{0.30\textwidth}

\usepackage{graphicx,verbatim}
\usepackage{dcolumn}
\usepackage{bm}
\graphicspath{{/user/rhaertle/epsfiles/Propplots/},{/user/rhaertle/Berlin08/},
{/user/rhaertle/epsfiles/Rate-Report/},{/user/rhaertle/epsfiles/RateCoul/},
{/user/rhaertle/epsfiles/Benesch/},{/user/rhaertle/PaperI/},{/user/rhaertle/PaperII/},
{/user/rhaertle/epsfiles/Cizek/},
{/user/rhaertle/epsfiles/PaperIV/},
{/user/rhaertle/PaperIII/}}

\begin{document}

\title{
Vibrational Instabilities in Resonant Electron Transport through Single-Molecule Junctions
}

\author{R. H\"artle}
\author{M. Thoss}
\affiliation{
Institut f\"ur Theoretische Physik und Interdisziplin\"ares Zentrum f\"ur Molekulare Materialien, \\
Friedrich-Alexander-Universit\"at Erlangen-N\"urnberg,\\ 
Staudtstr.\,7/B2, D-91058 Erlangen, Germany
}

\date{\today}

\begin{abstract}
We analyze various limits of vibrationally coupled resonant 
electron transport in single-molecule junctions. Based on a master
equation approach, 
we discuss analytic and numerical results
for junctions under a high bias voltage or weak electronic-vibrational coupling.
It is shown that in these limits the vibrational excitation of the molecular bridge increases indefinitely, 
\emph{i.e.}\ the junction exhibits a vibrational instability.
Moreover, our analysis 
provides analytic results for the vibrational distribution function and reveals that
these vibrational instabilities are related to electron-hole pair creation processes.
\end{abstract}

\maketitle

\section{Introduction}

Charge transport through nanostructures has been of great interest 
ever since nanofabrication techniques have been emerged \cite{Goldman1987,Leadbeater1989,Meir92,Davies93,Hyldgaard94,Kastner2000,Nitzan01,Selzer06,cuevasscheer2010}. 
The quantum mechanical nature of the charge-carriers in these nanostructures
gives rise to many intriguing transport phenomena, 
\emph{e.g.}\ strongly nonlinear transport characteristics \cite{Davies93,Hyldgaard94}. 
When these structures became continuously smaller, it was realized 
that vibrational (or phononic) degrees of freedom play an important role 
in this nonequilibrium transport problem \cite{Goldman1987,Leadbeater1989,Hyldgaard94,DAgosta06}. 
The ultimate limit of nanoelectronics 
is found in molecular electronic devices \cite{Nitzan01,Cuniberti05,Selzer06,Galperin07,cuevasscheer2010}, 
where the building blocks consist of single molecules. 
Due to their small size and mass, molecules often show strong
correlations between their electronic and vibrational degrees of freedom \cite{WuNazinHo04,LeRoy,Pasupathy05,Sapmaz06,Thijssen06,Parks07,Boehler07,Leon2008,Huettel2009,Tao2010,Ballmann2010,Jewell2010,Osorio2010}.
It is thus of great interest to understand electron transport through a nanostructure 
that exhibits electronic-vibrational coupling,
such as a single molecule coupled to a left and a right electrode \cite{Galperin07,Hartle2010b}.
Analyzing the limits of this transport problem, especially in the resonant transport regime, facilitates the understanding of nonequilibrium transport at the nanoscale.

\begin{figure}
\begin{tabular}{cccc}
\resizebox{\newwidthprime}{\newheightprime}{
\includegraphics{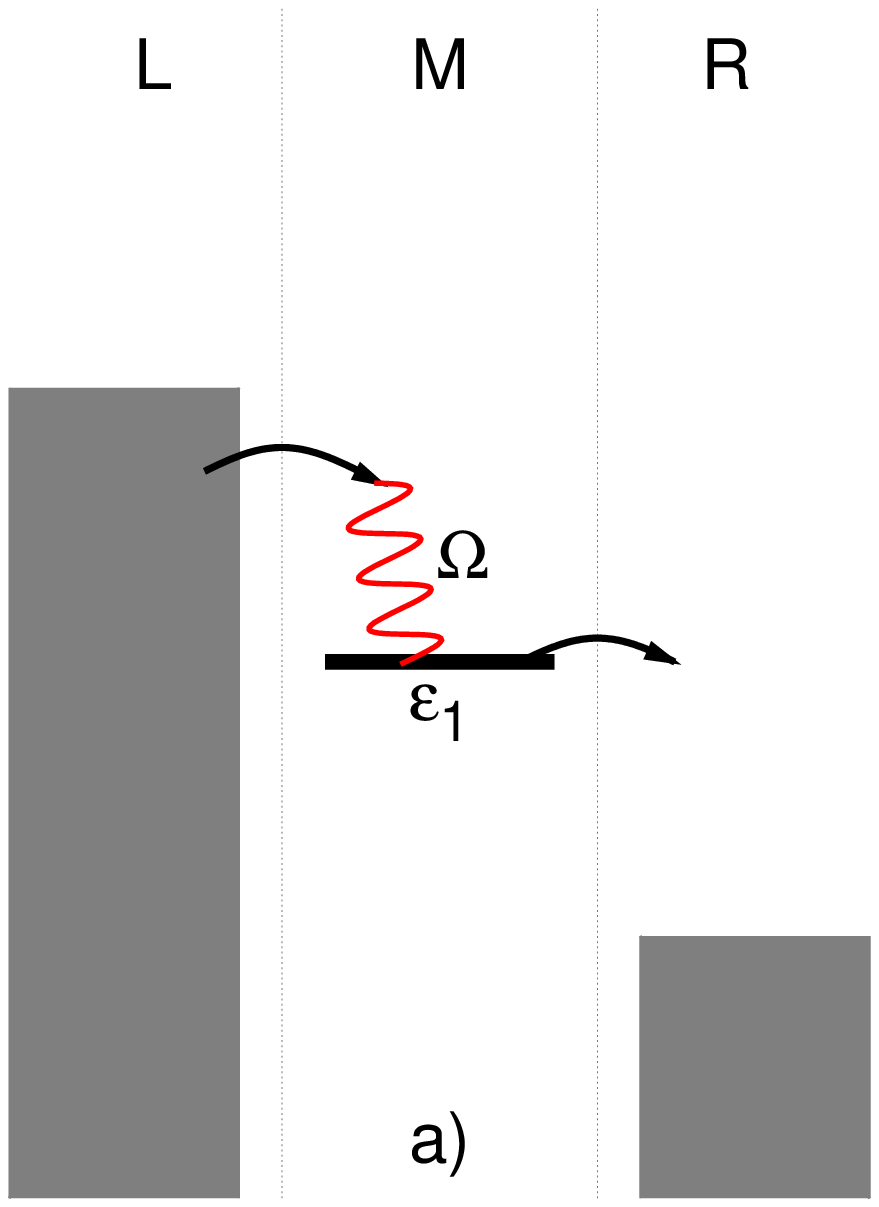}
}
&
\resizebox{\newwidthprime}{\newheightprime}{
\includegraphics{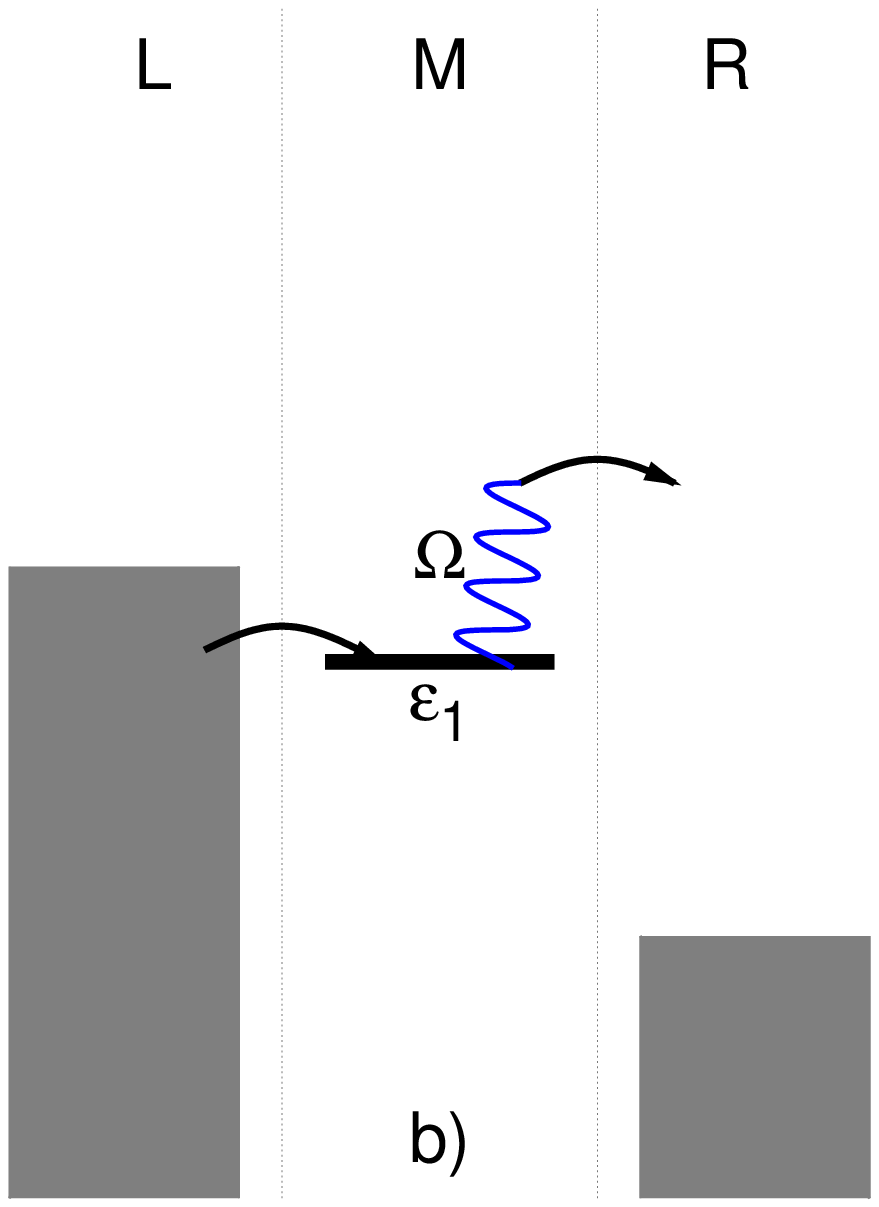}
}
&
\resizebox{\newwidthprime}{\newheightprime}{
\includegraphics{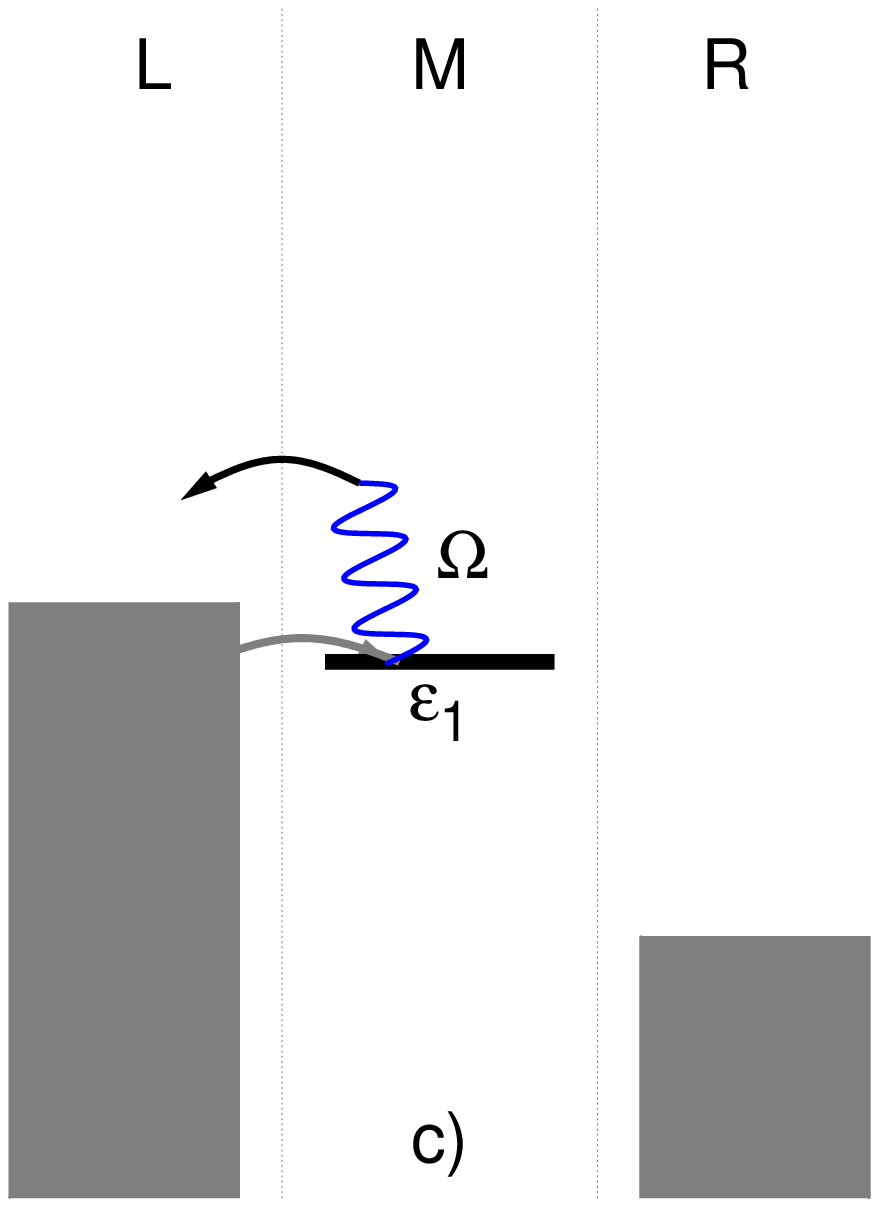}
}
&
\\ 
\end{tabular}
\caption{\label{simpleemissionandabsorption} Basic processes in vibrationally coupled electron transport through single-molecule junctions. 
Panel a) (Panel b)) depicts an example for an emission (absorption) process, where an electron sequentially tunnels from the left lead onto the molecule and further to the right lead, thereby singly exciting (deexciting) the vibrational mode of the molecular bridge (red (blue) wiggly line). The frequency of the vibrational mode is denoted by $\Omega$. An example for a process, where an electron-hole pair is created in the left lead by singly deexciting the vibrational mode of the bridge, is shown in Panel c).}
\end{figure}

Due to electronic-vibrational coupling, transport through a single-molecule junction comprises not only charge-exchange but also energy-exchange processes with the leads. Examples of such processes are depicted in Fig.\ \ref{simpleemissionandabsorption}. Thereby, Panel a) and b) represent transport processes, where the tunneling electron vibrationally excites (heating) and deexcites (cooling) the molecular bridge, respectively. Panel c) shows a process, where in two sequential tunneling processes an electron-hole pair is created in the left lead upon absorption of vibrational energy from the molecular bridge.
Such energy-exchange processes with the leads play an important role in molecular junctions. 
This can be illustrated by considering a junction, 
where the temperature in the leads, $T$, represents the largest energy scale. 
In this limit the population of the molecular energy levels is 
determined by the thermal distribution that also determines the 
population of the levels in the leads. 
The corresponding average vibrational excitation of the molecular bridge thus increases 
indefinitely as $T\rightarrow\infty$. 
That way, the high-temperature limit exhibits a vibrational instability in a trivial sense. 
Similarly, the static limit \cite{Galperin2008}, where the frequency of the vibrational modes represent the smallest energy scale, $\Omega\rightarrow0$ ($k_{\text{B}}T\gg\Omega$), results in an infinite vibrational excitation, since 
energy-exchange processes transfer the thermal excitation of the leads to 
the vibrational degrees of freedom of the molecular bridge. 

In this article, we focus on the limit of a large bias voltage, $\Phi\rightarrow\infty$, 
as well as a weak electronic-vibrational coupling, $\lambda\rightarrow0$, 
and analyze the vibrational instabilities inherent to these limits. 
For an infinite bias voltage we find the corresponding vibrational excitation to diverge, 
because electron-hole pair creation processes are completely suppressed. 
This important cooling mechanism is also missing for vanishing electronic-vibrational coupling, $\lambda\rightarrow0$, 
if the bias voltage is large enough. As a result, 
the limit of weak electronic-vibrational coupling exhibits another vibrational instability. 
This intriguing phenomenon was already reported and analyzed by Koch \emph{et al.} \cite{Semmelhack}. 
Resonant absorption processes with a higher-lying electronic state may prevent this instability \cite{Hartle09}. 
For moderate bias voltages that allow for the leading-order electron-hole pair 
creation processes in the limit $\lambda\rightarrow0$, we derive an analytic expression 
for the respective vibrational distribution function, 
which shows that the corresponding average vibrational excitation is finite and nonzero. 
Only in the off-resonant transport regime, \emph{i.e.}\ for small bias voltages, we find that vibrational excitation 
vanishes as $\lambda\rightarrow0$.

To study these limits we employ a master equation formalism \cite{May02,Mitra04,Lehmann04,Pedersen05,Harbola2006,Zazunov06,Siddiqui,Timm08,May08,May08b,Leijnse09,Esposito09,Esposito2010}
that is based on a second-order expansion in the coupling of the molecular bridge to the leads.  
This allows a description of all resonant transport processes, but misses higher-order processes such as the broadening of levels due to the 
coupling between the molecule and the leads or equivalently co-tunneling processes \cite{Galperin2007,Lueffe,Hartle,Leijnse2009}.
Such processes can, in principle, be described by approaches that take into account higher-order effects, 
including advanced master equation approaches \cite{May02,Pedersen05,Leijnse09,Esposito09,Esposito2010,Timm2010}, 
scattering theory approaches \cite{Cizek04,Toroker07,Zimbovskaya09,Seidemann10}, 
nonequilibrium Green's function methods 
\cite{Flensberg03,Mitra04,Galperin06,Ryndyk06,Frederikesen07b,Tahir2008,Hartle,Stafford09,Haupt2009,Hartle09,Wohlman2010}, or 
numerically exact methodologies \cite{Muehlbacher08,Thorwart2008,Wang09,Hackl10,Segal2010}. 
However, since we are primarily interested in the limits of resonant electron transport 
through single-molecule junctions, higher-order processes play no role for our considerations.

The article is organized as follows. In Section \ref{theorySec}, we introduce the Hamiltonian (Sec.\ \ref{modelHamSec}) and the master equation approach (Sec.\ \ref{theory}) that we use to describe vibrationally coupled electron transport through single-molecule junctions in the resonant transport regime. We present our results in Sec.\ \ref{results}, wherein the transport characteristics of a molecular junction are analyzed in the limit of an infinite bias voltage (Sec.\ \ref{highbiasSec}) and for a vanishing electronic-vibrational coupling (Sec.\ \ref{lambdazero}). The latter case is studied in three different regimes: in Sec.\ \ref{vibrinstab} for a high bias voltage, where electron-hole pair creation processes are suppressed, in Sec.\ \ref{nonanalytic} for a lower bias voltage that allows for electron-hole pair creation processes, and in Sec.\ \ref{higherstate} including a second higher-lying electronic state that is vibrationally coupled.

\section{Theory}
\label{theorySec}

\subsection{Model Hamiltonian}
\label{modelHamSec}

We investigate vibrationally coupled electron
transport through single-molecule junctions
using the following Hamiltonian (throughout the article we use units where $\hbar=1$)
\begin{eqnarray}
H &=& \sum_{i\in\text{M}} \epsilon_{i} c_{i}^{\dagger}c_{i} + 
\sum_{k\in\text{L,R}} \epsilon_{k} c_{k}^{\dagger}c_{k} + \sum_{k,i} ( V_{ki} 
c^{\dagger}_{k} c_{i} + \text{h.c.} ) \nonumber\\
&&\hspace{-0cm} + \Omega
a^{\dagger}a + \sum_{i\in\text{M}} \lambda_{i} 
(a+a^{\dagger}) \left( c_{i}^{\dagger}c_{i} - \delta_{i} \right),
\end{eqnarray}
where electronic states with energies $\epsilon_{i}$,
located at the molecular bridge (M), 
are coupled by interaction 
matrix elements $V_{ki}$ to electronic states in the
leads (L,R). The energies of the lead states are labelled by $\epsilon_{k}$.
The operators $c^{\dagger}_{i}$/$c_{i}$ ($c^{\dagger}_{k}$/$c_{k}$) denote respective creation/annihilation operators for the states of the molecular bridge (leads).
To model the vibrational degrees of freedom of the junction we 
consider a single harmonic mode with frequency $\Omega$ that is
described by creation/annihilation operators $a^{\dagger}$/$a$.
The coupling strength between the harmonic mode
and the $i$th state of the molecular bridge is denoted by $\lambda_{i}$. 
Thereby, the fixed parameters $\delta_{i}$ ensure that 
there is no electronic-vibrational coupling in the electronic ground-state of the neutral molecular junction \cite{Cederbaum74,Benesch08,Hartle2010b}. 
This means that $\delta_{i}=0$ ($\delta_{i}=1$), if the $i$th electronic state is unoccupied (occupied) in this reference state.

We prediagonalize $H$ by the small polaron transformation \cite{Mitra04},
 $H\rightarrow\overline{H} = \overline{H}_{\text{S}} + \overline{H}_{\text{B}} + \overline{H}_{\text{SB}}$,
resulting in
\begin{eqnarray}
 \overline{H}_{\text{S}} &=& \sum_{i} \overline{\epsilon}_{i} c_{i}^{\dagger}c_{i}  - 2 \sum_{i<j} \frac{\lambda_{i}\lambda_{j}}{\Omega} \left( c_{i}^{\dagger}c_{i} -\delta_{i} \right) \left( c_{j}^{\dagger}c_{j} -\delta_{j} \right)
  + \Omega a^{\dagger} a,  \\
 \overline{H}_{\text{B}} &=& \sum_{k} \epsilon_{k} c_{k}^{\dagger}c_{k}, \\ \overline{H}_{\text{SB}} &=& \sum_{ki} ( V_{ki} X_{i} 
c_{k}^{\dagger}c_{i} + \text{h.c.} ). 
\end{eqnarray}
Thereby, $\overline{H}_{\text{S}}$ comprises the degrees of freedom of the molecular bridge (S), 
in particular the polaron-shifted electronic states, 
$\overline{\epsilon}_{i}=\epsilon_{i}+(2\delta_{i}-1)(\lambda_{i}^{2}/\Omega)$, 
vibrationally induced electron-electron interactions, $\sim 2\lambda_{i}\lambda_{j}/\Omega $,
and the harmonic mode. 
The leads degrees of freedom (B) are 
summarized in $\overline{H}_{\text{B}}$. 
The term $\overline{H}_{\text{SB}}$
describes the coupling between the molecule and the leads,
which, as a result of electronic-vibrational coupling, 
is renormalized by shift operators 
$X_{i}=\text{exp}((\lambda_{i}/\Omega)(a-a^{\dagger}))$.
Due to the small polaron transformation, there is no explicit electronic-vibrational coupling term in $\overline{H}_{\text{S}}$. 

\subsection{Transport Theory}

\label{theory}

The 
current and vibrational excitation 
of a single-molecule junction 
are obtained from the reduced density 
matrix $\rho$, which is given as the stationary limit of the well-established 
equation of motion \cite{May02,Mitra04,Lehmann04,Harbola2006,Volkovich2008,Hartle09,Hartle2010,Hartle2010b}
\begin{eqnarray}
\label{genfinalME}
\frac{\partial \rho(t)}{\partial t} &=& -i \left[ \overline{H}_{\text{S}} , \rho(t) \right]  - \int_{0}^{\infty} \text{d}\tau\, \text{tr}_{\text{B}}\lbrace \left[ \overline{H}_{\text{SB}} , \left[ \overline{H}_{\text{SB}}(\tau), \rho(t) \rho_{\text{B}} \right] \right] \rbrace , 
\end{eqnarray}
with 
\begin{eqnarray}
\overline{H}_{\text{SB}}(\tau)=\text{e}^{-i(\overline{H}_{\text{S}}+\overline{H}_{\text{B}})\tau} \overline{H}_{\text{SB}} \text{e}^{i(\overline{H}_{\text{S}}+\overline{H}_{\text{B}})\tau}.
\end{eqnarray}
Here, $\rho_{\text{B}}$ represents the equilibrium density matrix of the leads. Eq.\ (\ref{genfinalME}) can be derived from the Nakajima-Zwanzig equation \cite{Nakajima,Zwanzig}, 
employing a second-order expansion in the coupling $\overline{H}_{\text{SB}}$ along with the so-called Markov approximation
\begin{eqnarray}
\rho(t-\tau) \approx \text{e}^{i\overline{H}_{\text{S}}\tau} \rho(t) \text{e}^{-i\overline{H}_{\text{S}}\tau}.
\end{eqnarray}
The master equation, Eq.\ (\ref{genfinalME}), is
evaluated assuming a stationary state, $\frac{\partial \rho}{\partial t}=0$, 
and with 
basis functions $\vert a \rangle\vert \nu \rangle$ 
that span the subspace of the electronic 
$\vert a \rangle$ and the vibrational  degrees of freedom $\vert \nu \rangle$, respectively. 
Thereby, the electronic basis functions are given in the occupation number 
representation \emph{i.e.}\ $\vert a \rangle=\vert n_{1}n_{2}.. \rangle$, where 
$n_{i}\in\lbrace0,1\rbrace$ denotes the population of the $i$th electronic state. 
The vibrational basis function $\vert\nu\rangle$ with 
$\nu\in\mathbb{N}_{0}$ represents the $\nu$th level of the harmonic mode. 
The coefficients of the reduced density matrix are thus denoted by
\begin{eqnarray}
 \rho_{a,a'}^{\nu_{1}\nu_{2}} \equiv \langle a \vert \rho^{\nu_{1}\nu_{2}} 
\vert a' \rangle \equiv \langle a \vert \langle \nu_{1} \vert \rho \vert \nu_{2} 
\rangle \vert a' \rangle.
\end{eqnarray}
In the evaluation of Eq.\ (\ref{genfinalME}), we neglect principal value terms that describe the renormalization of 
the molecular energy levels due to the coupling between the bridge and the leads \cite{Harbola2006,Hartle2010b}. These contributions are irrelevant for the results discussed in this work.

Having determined the coefficients of the reduced density matrix, 
we can readily obtain the vibrational distribution function
\begin{eqnarray}
 p_{\nu} &=& \sum_{a} \rho^{\nu\nu}_{a,a} .
\end{eqnarray}
The corresponding average vibrational excitation 
for a molecular junction with a single electronic state
is given by
\begin{eqnarray}
\langle a^{\dagger}a \rangle_{H} &=& \langle a^{\dagger} a \rangle_{\overline{H}} + 
\frac{\lambda^{2}}{\Omega^{2}} ( n_{1} - 2 \delta_{1} n_{1} + \delta_{1} ), \\
&=&   \sum_{\nu} \nu p_{\nu}+ 
\frac{\lambda^{2}}{\Omega^{2}} ( n_{1} - 2 \delta_{1} n_{1} + \delta_{1} ), \nonumber
\end{eqnarray}
with 
\begin{eqnarray}
 n_{1} = \langle c^{\dagger}_{1} c_{1} \rangle_{H} =
\sum_{\nu} \rho^{\nu\nu}_{1,1}.
\end{eqnarray}
For transport through a molecular junction with two electronic states, 
the average vibrational excitation is calculated according to
\begin{eqnarray}
\langle a^{\dagger}a \rangle_{H} &=&  \sum_{\nu} \nu p_{\nu} + 
\frac{\lambda_{1}^{2}}{\Omega^{2}} ( n_{1} - 2 \delta_{1} n_{1} + \delta_{1} )  + \frac{\lambda_{2}^{2}}{\Omega^{2}} ( n_{2} - 2 \delta_{2} n_{2} + \delta_{2} ) \\
&&  + 
2\frac{\lambda_{1}\lambda_{2}}{\Omega^{2}} ( \sum_{\nu} \rho^{\nu\nu}_{11,11} -\delta_{2} n_{1} - \delta_{1} n_{2} + \delta_{1}\delta_{2} ) , \nonumber
\end{eqnarray}
with  
\begin{eqnarray}
 n_{1} &=& \langle c^{\dagger}_{1} c_{1} \rangle_{H} =  \sum_{\nu} \rho^{\nu\nu}_{11,11} 
+ \rho^{\nu\nu}_{10,10},\\
n_{2} &=& \langle c^{\dagger}_{2} c_{2} \rangle_{H} =  \sum_{\nu} \rho^{\nu\nu}_{11,11} 
+ \rho^{\nu\nu}_{01,01}.
\end{eqnarray}

The current through the molecular junction
is obtained from the formula
\begin{eqnarray}
\label{currentgen}
I=-i\int_{0}^{\infty}\text{d}\tau\, \text{tr}_{\text{S+B}}\lbrace \left[\overline{H}_{\text{SB}}(\tau),\rho \rho_{\text{B}}\right] \hat{I}\rbrace,
\end{eqnarray}
with
$ \hat{I} = -2e\frac{\text{d}}{\text{d}t} \sum_{k\in\text{L}}  c_{k}^{\dagger}c_{k}$. For the derivation 
and evaluation of Eq.\ (\ref{currentgen}), the same approximations as for Eq.\ (\ref{genfinalME}) have been used.

\section{Results}

\label{results}

In this section, we use the methodology outlined in Sec.\ \ref{theory} to 
investigate various limits of resonant electron transport through single molecules. 
In particular, we study the current and the vibrational excitation of a molecular junction 
in the limit of an infinite bias voltage (Sec.\ \ref{highbiasSec}) and for vanishing vibronic coupling (Sec.\ \ref{lambdazero}).
Our analysis includes analytic as well as numerical results. For the latter we have used a set of $400$ 
vibrational basis functions to obtain numerical convergence.
Since we consider systems without quasi-degeneracies, coherences of the density matrix are negligible, \emph{i.e.}\ $\rho^{\nu_{1}\nu_{2}}_{a,a'}= \rho^{\nu_{1}\nu_{1}}_{a,a} \delta_{\nu_{1}\nu_{2}}\delta_{a,a'}$.
Furthermore, we use the wide-band approximation and assume the bias voltage $\Phi$ to drop symmetrically at the contacts.

\subsection{The high-bias limit $\Phi\rightarrow\infty$}
\label{highbiasSec}

\begin{figure}
\resizebox{\newwidth}{\newheight}{
\includegraphics{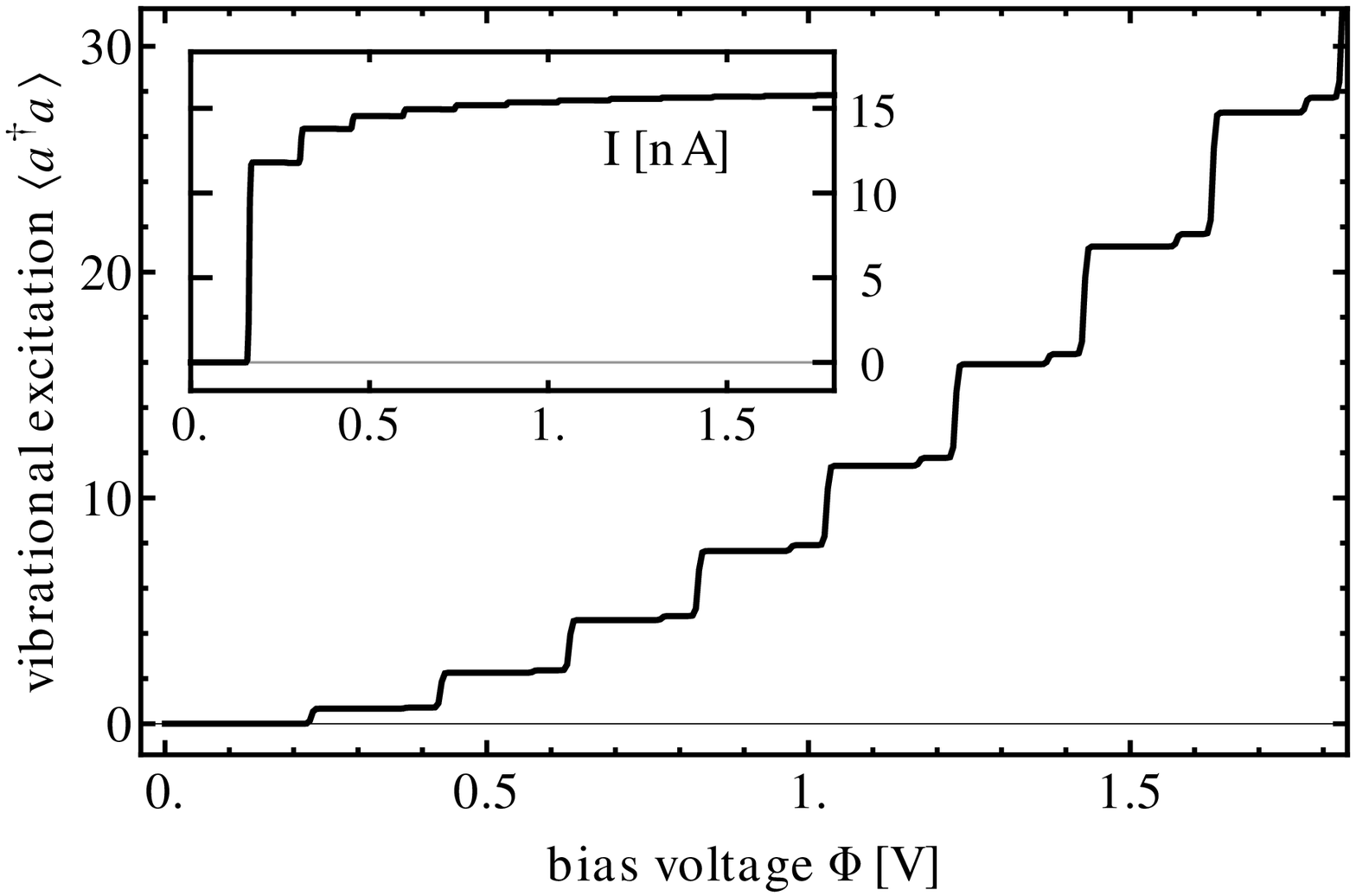}
}
\resizebox{\newwidth}{\newheight}{
\includegraphics{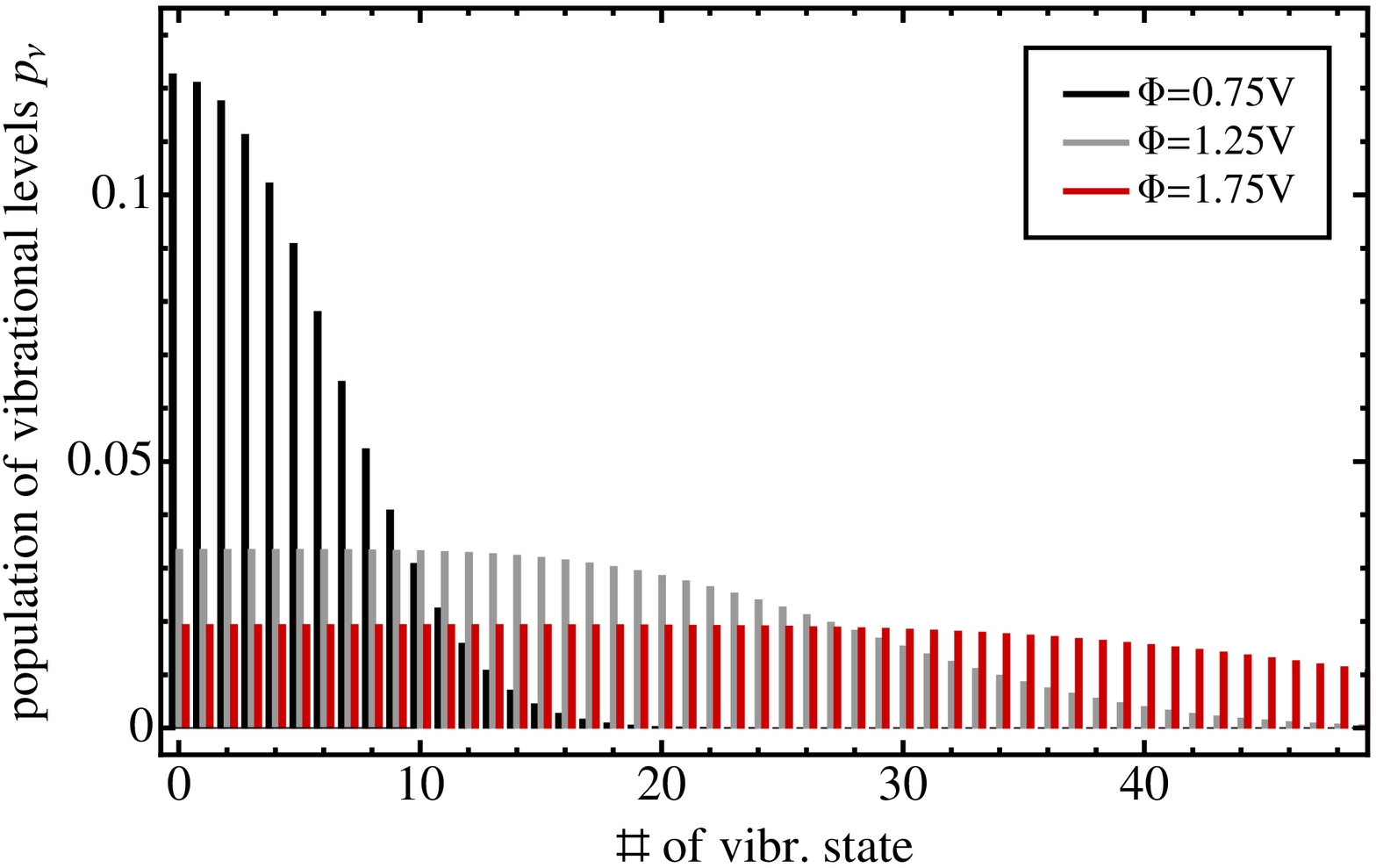}
}
\caption{\label{highbias} (Color online) \emph{Upper Panel}: Average vibrational excitation, $\langle a^{\dagger}a \rangle$, of a molecular junction with a single electronic state that is coupled to a single vibrational mode. The inset shows the respective current-voltage characteristics.
\emph{Lower Panel}: Population of vibrational levels, $p_{\nu}$, corresponding to three 
different values of the bias voltage $\Phi$. 
Increasing the bias voltage, the vibrational distribution function becomes broader, and thus, the level of vibrational excitation gets higher. In the limit $\Phi\rightarrow\infty$ this leads to an infinite vibrational excitation.}
\end{figure}

In the resonant transport regime, the average vibrational excitation of a single-molecule junction typically increases very rapidly with increasing bias voltage \cite{Hartle,Hartle09,Hartle2010b}. An example for this behavior is given in Fig.\ \ref{highbias}a, which shows the vibrational excitation of a molecular junction induced by inelastic transport processes through a single electronic state (like the one depicted in Fig.\ \ref{simpleemissionandabsorption}a). 
Increasing the bias voltage, more and more inelastic transport processes become active. However, 
as they also involve an increasing number of vibrational quanta, which results in an unfavorable Franck-Condon overlap,
they are typically strongly suppressed. 
Hence, the question arises wether the level of vibrational excitation, 
induced by these processes, does or does not saturate in the limit $\Phi\rightarrow\infty$.

To prove that vibrational excitation increases indefinitely in the limit $\Phi\rightarrow\infty$, we assume in the following that vibrational excitation is finite and derive a contradiction to this assumption.
To this end, we evaluate Eq.\ (\ref{genfinalME}) between the basis functions $\langle 0 \vert\langle \nu_{1} \vert$ and $\vert \nu_{1} \rangle\vert 0 \rangle$,
\begin{eqnarray}
&&  \sum_{K\nu_{3}}   f_{K}(\overline{\epsilon}_{1}  + \Omega(\nu_{3}-\nu_{1}))   \Gamma_{K} \vert  X_{1,\nu_{1}\nu_{3}}  \vert^{2} \rho_{0,0}^{\nu_{1}\nu_{1}}  = \qquad\\
&&  \sum_{K\nu_{3}}   (1-f_{K}(\overline{\epsilon}_{1}  + \Omega(\nu_{3}-\nu_{1})))  \Gamma_{K}  \vert  X_{1,\nu_{1}\nu_{3}}  \vert^{2}  \rho^{\nu_{3}\nu_{3}}_{1,1}.   \nonumber
\end{eqnarray}
Multiplying this equation by $\nu_{1}$, and summing up all equations, we obtain the following equation
\begin{eqnarray}
\label{highbias3}
&&  \sum_{K\nu_{1}\nu_{3}}  \nu_{1} f_{K}(\overline{\epsilon}_{1}  + \Omega(\nu_{3}-\nu_{1}))   \Gamma_{K} \vert  X_{1,\nu_{1}\nu_{3}}  \vert^{2} \rho_{0,0}^{\nu_{1}\nu_{1}}  = \qquad\\
&&  \sum_{K\nu_{1}\nu_{3}}  \nu_{1} (1-f_{K}(\overline{\epsilon}_{1}  + \Omega(\nu_{3}-\nu_{1})))  \Gamma_{K}  \vert  X_{1,\nu_{1}\nu_{3}}  \vert^{2}  \rho^{\nu_{3}\nu_{3}}_{1,1}  . \nonumber
\end{eqnarray}
If we assume vibrational excitation $\langle a^{\dagger}a \rangle$  to converge in the limit $\Phi\rightarrow\infty$, so must the \emph{lhs} and the \emph{rhs} of the above equation. Hence, we can take the limit $\Phi\rightarrow\infty$ on both sides in Eq.\ (\ref{highbias3}), replacing $f_{\text{L}}(\epsilon)$ by $1$ and $f_{\text{R}}(\epsilon)$ by $0$, which results in
\begin{eqnarray}
\sum_{\nu_{1}\nu_{3}}  \nu_{1} \Gamma_{\text{L}} \vert  X_{1,\nu_{1}\nu_{3}}  \vert^{2} \rho_{0,0}^{\nu_{1}\nu_{1}}  =  \sum_{\nu_{1}\nu_{3}}  \nu_{1} \Gamma_{\text{R}}  \vert  X_{1,\nu_{1}\nu_{3}}  \vert^{2}  \rho^{\nu_{3}\nu_{3}}_{1,1}  .
\end{eqnarray}
Applying the sum rule $\sum_{\nu}\vert X_{\nu\mu}\vert^{2}=1$ to the \emph{lhs}, and  $\sum_{\nu}\nu\vert X_{\nu\mu}\vert^{2}=\lambda^{2}/\Omega^{2}+\mu$ to the \emph{rhs} gives
\begin{eqnarray}
\label{highbias4}
\sum_{\nu_{1}}  \nu_{1} \Gamma_{\text{L}} \rho_{0,0}^{\nu_{1}\nu_{1}}  &=&  \sum_{\nu_{3}}  (\nu_{3}+\lambda^{2}/\Omega^{2}) \Gamma_{\text{R}}  \rho^{\nu_{3}\nu_{3}}_{1,1} . 
\end{eqnarray}
Analogously, using the $\langle 1 \vert\langle \nu_{1} \vert ... \vert \nu_{1} \rangle\vert 1 \rangle$-projection of Eq.\ (\ref{genfinalME}), we obtain:
\begin{eqnarray}
\label{highbias5}
  \sum_{\nu_{3}}  (\nu_{3}+\lambda^{2}/\Omega^{2}) \Gamma_{\text{L}}  \rho^{\nu_{3}\nu_{3}}_{0,0}  &=& \sum_{\nu_{1}}  \nu_{1} \Gamma_{\text{R}} \rho_{1,1}^{\nu_{1}\nu_{1}} .
\end{eqnarray}
Subtracting Eq.\ (\ref{highbias5}) from Eq.\ (\ref{highbias4}) leads to the following equation
\begin{eqnarray}
\label{contra}
- \lambda^{2}/\Omega^{2} \sum_{\nu_{3}}  \Gamma_{\text{L}}  \rho^{\nu_{3}\nu_{3}}_{0,0}   &=&  \lambda^{2}/\Omega^{2} \sum_{\nu_{3}}  \Gamma_{\text{R}}  \rho^{\nu_{3}\nu_{3}}_{1,1},   
\end{eqnarray}
which is a contradiction, since the \emph{lhs} of Eq.\ (\ref{contra}) is negative while its \emph{rhs} is positive, if $\lambda\neq0$. Hence, for finite electronic-vibrational coupling $\lambda$, 
vibrational excitation must diverge as $\Phi\rightarrow\infty$.

Fig.\ \ref{highbias} shows numerical results 
that illustrate this analytic finding. For these calculations, we have employed 
a model for a molecular junction that consists of a single electronic state at $\overline{\epsilon}_{1}=0.15$\,eV and a single vibrational mode with frequency $\Omega=0.1$\,eV. The corresponding electronic-vibrational coupling strength is $\lambda=0.06$\,eV. 
These parameters represent typical values for molecular junctions similar to those 
that have been employed in first-principles models \cite{Pecchia04,Frederiksen04,Benesch06,Troisi2006,Benesch08,Benesch2009,Monturet2010}. 
The level-width functions, $\Gamma_{\text{L}}=\Gamma_{\text{R}}=67$\,$\mu$eV, are chosen to be much smaller than the thermal broadening $k_{\text{B}}T=1$\,meV \cite{noteonMoleculeLeadCoupling}. 
The black, gray and red bars in Fig.\ \ref{highbias}b represent the corresponding vibrational distribution function for 
bias voltages $\Phi=0.75$\,V, $1.25$\,V and $1.75$\,V, respectively. It is seen that the vibrational distribution function becomes broader with increasing bias voltage and exhibits equal population of the lower vibrational states for large bias voltages. 
This leads to a continuous increase in vibrational excitation with increasing bias voltage, which does not saturate. 

We attribute this behavior to the lack of electron-hole pair creation processes (cf.\ Fig.\ \ref{simpleemissionandabsorption}c), which for $\Omega\gg k_{\text{B}}T$ can only deexcite the vibrational mode \cite{Hartle2010b}, and which are completely blocked in the high bias limit. 
As a consequence, the number of excitation and deexcitation processes upon electron transport through the molecule is equal. 
The respective stationary state is given by a vibrational distribution function, where all vibrational levels are equally populated (cf.\ Fig.\ \ref{highbias}b). This can be rationalized considering the population of the $\nu$th vibrational level $p_{\nu}$ after any transport process in this limit. Since there are as many excitation as deexcitation processes, the respective population is given by the sum $\sum_{\nu'=1}^{\infty} \tilde{p}_{\nu'} \vert X_{\nu\nu'}\vert^{2}$ with $\tilde{p}_{\nu}$ the corresponding population before such a transport process. The only nonequilibrium state, which is invariant under these conditions, \emph{i.e.}\ $p_{\nu}=\tilde{p}_{\nu}$, is that, where all vibrational levels are equally occupied since $\sum_{\nu'=1}^{\infty} \vert X_{\nu\nu'}\vert^{2}=1$.

The respective current is given by $2e\Gamma_{\text{L}}\Gamma_{\text{R}}/(\Gamma_{\text{L}}+\Gamma_{\text{R}})$, which is the same result as Gurvitz \emph{et al.}\ \cite{Gurvitz96} found for a junction in the high-bias limit without electronic-vibrational coupling.

\subsection{Weak electronic-vibrational coupling $\lambda\rightarrow0$}
\label{lambdazero}

For a finite bias voltage that fulfills $e\Phi>2(\overline{\epsilon}_{1}+\Omega)$, 
the vibrational excitation may also diverge in the limit 
of vanishing vibronic coupling $\lambda\rightarrow0$. This counter-intuitive phenomenon was reported before \cite{Mitra04,Semmelhack}. In subsection \ref{vibrinstab}, we give a short overview of the phenomenon, 
and reinterpret the phenomenon in terms of electron-hole pair creation processes.
This way, we establish the relation between the limit of a weak vibronic coupling and the high-bias limit 
discussed in the previous section.
In the two subsequent sections, we investigate the limit $\lambda\rightarrow0$ for lower bias voltages (Sec.\ \ref{nonanalytic}), where the leading-order electron-hole pair creation processes are not blocked, and in the presence of a second higher-lying electronic state that is vibrationally coupled (Sec.\ \ref{higherstate}). 

\subsubsection{Without cooling by electron-hole pair creation processes}
\label{vibrinstab}

\begin{figure}
\resizebox{\newwidth}{\newheight}{
\includegraphics{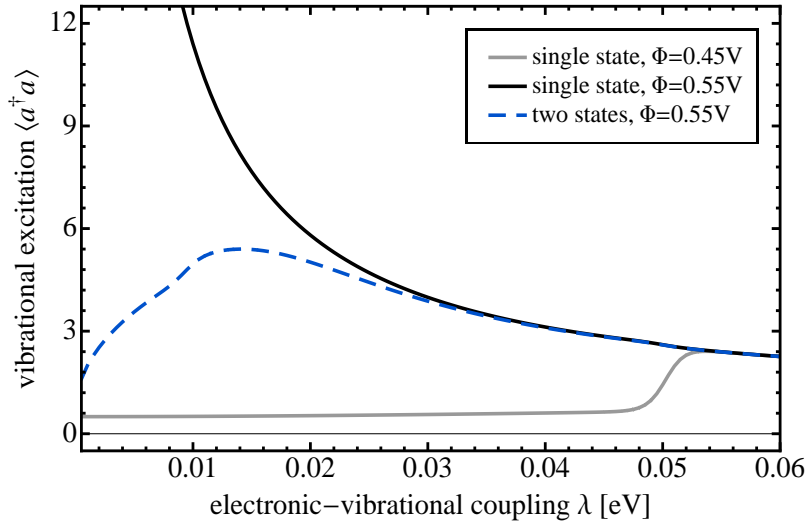}
}
\caption{\label{cureofvibrinst} (Color online) Vibrational excitation as a function of the electronic-vibrational coupling strength $\lambda$. The gray line shows the vibrational excitation obtained for the model system with a single electronic state at a bias voltage $2(\overline{\epsilon}_{1}+\Omega)>e\Phi>2\overline{\epsilon}_{1}$, reflecting the result of Sec.\ \ref{nonanalytic}. The black line represents the phenomenon of vibrational instability, which is outlined in Sec.\ \ref{vibrinstab}. The dashed blue line shows that a higher-lying electronic state, which is coupled to the vibrational mode, prevents the vibrational instability depicted by the black line (cf.\ Sec.\ \ref{higherstate}).}
\end{figure}

The solid black line in Fig.\ \ref{cureofvibrinst} illustrates the phenomenon of vibrational instability in the limit $\lambda\rightarrow0$. 
It represents the vibrational excitation, resulting from vibrationally coupled transport through a single electronic state, as a function of the electronic-vibrational coupling strength $\lambda$ at a fixed bias voltage $e\Phi=0.55$\,eV $>2(\overline{\epsilon}_{1}+\Omega)$. The model parameters are the same as those described in Sec.\ \ref{highbiasSec}. Upon reducing $\lambda$, the level of vibrational excitation monotonously increases \cite{Mitra04,Semmelhack,Hartle09,Hartle2010b}, where the slope of this increase gets larger for smaller values of $\lambda$. Moreover, the vibrational distribution function (data not shown) becomes broader and shows an increasing number of vibrational levels, starting from the ground-state, that are equally populated. 
Interestingly, this phenomenon occurs only if the associated bias voltage $e\Phi$ exceeds $2(\overline{\epsilon}_{1}+\Omega)$. For such bias voltages electron-hole pair creation processes (cf.\ Fig.\ \ref{simpleemissionandabsorption}c) are suppressed to lowest order in $\lambda$. In this sense, the limit $\lambda\rightarrow0$ is equivalent to the high-bias limit, where also all relevant electron-hole pair creation processes become blocked and the level of vibrational excitation diverges for $\Phi\rightarrow\infty$.

These numerical findings can only give an example of the phenomenon, and are, furthermore, 
limited by the number of basis functions employed. In the following, 
we therefore analyze this behavior for $\lambda\rightarrow0$ in an analytic and more general way.
To second order in $\lambda$, with $e\Phi>2(\overline{\epsilon}_{1}+\Omega)>2\Omega$, we obtain from Eq.\ (\ref{genfinalME}) the following set of equations ($\nu\geq1$)
\begin{eqnarray}
\label{vibrinstab-one}
 0 &=&  \Gamma_{\text{L}} \rho_{0,0}^{\nu\nu} -   \Gamma_{\text{R}}  \rho^{\nu\nu}_{1,1}   -   \Gamma_{\text{R}} \left(  \nu \lambda^{2}  \rho^{\nu-1\nu-1}_{1,1} - (2\nu+1) \lambda^{2} \rho^{\nu\nu}_{1,1} +  (\nu+1)\lambda^{2} \rho^{\nu+1\nu+1}_{1,1}  \right), \\
\label{vibrinstab-two}
 0 &=&  \Gamma_{\text{L}} \rho_{0,0}^{\nu\nu} -   \Gamma_{\text{R}}  \rho^{\nu\nu}_{1,1}   +   \Gamma_{\text{L}} \left(  \nu \lambda^{2}  \rho^{\nu-1\nu-1}_{0,0} - (2\nu+1) \lambda^{2} \rho^{\nu\nu}_{0,0} +  (\nu+1)\lambda^{2} \rho^{\nu+1\nu+1}_{0,0}  \right). 
\end{eqnarray}
From the latter equations one infers that the difference $\Gamma_{\text{L}} \rho_{0,0}^{\nu\nu} -   \Gamma_{\text{R}}  \rho^{\nu\nu}_{1,1}$ is of second order in $\lambda$. Therefore, we can
replace the terms $\Gamma_{\text{R}}\rho^{\nu\nu}_{1,1}$ by $\Gamma_{\text{L}}\rho^{\nu\nu}_{0,0}$ in Eq.\ (\ref{vibrinstab-one}) (or vice versa in Eq.\ (\ref{vibrinstab-two})). 
Subtracting Eq.\ (\ref{vibrinstab-one}) from Eq.\ (\ref{vibrinstab-two}) (or vice versa) thus gives
\begin{eqnarray}
\label{vibrinseries}
 0 &=& (2\nu+1) \rho_{a,a}^{\nu\nu} - (\nu+1)\rho_{a,a}^{\nu+1\nu+1} - \nu \rho_{a,a}^{\nu-1\nu-1} ,
\end{eqnarray}
where $a\in\{0,1\}$. The recurrence relation defined by Eq.\ (\ref{vibrinseries}) leads to divergent populations $\rho_{a,a}^{\nu\nu}\stackrel{\nu\rightarrow\infty}{\rightarrow}\pm\infty$, if $\rho_{a,a}^{11}-\rho_{a,a}^{00}\neq0$. 
Thus, the only solution, which is normalizable, is the one, where all vibrational levels are equally occupied, $\rho_{a,a}^{\nu+1\nu+1}-\rho_{a,a}^{\nu\nu}=0$. This corresponds to an infinite vibrational excitation or a vibrational instability in the limit $\lambda\rightarrow0$. 

It is noted that rigorous divergence of the vibrational excitation is only found for an isolated molecular vibration,
as it is described by the standard model of vibrationally coupled electron transport in molecular junctions considered here.
In real molecular junctions, vibrational relaxation processes, introduced 
\emph{e.g.}\ by coupling to phonons of the electrodes or other vibrational modes, 
would restrict the vibrational excitation to a finite value. 
However, even in the presence of such relaxation mechanisms, vibrational excitation 
may not only monotonously increase with $\lambda$, but may also decrease with an 
increasing electronic-vibrational coupling strength.

As in the high-bias limit, the current obtained for $\lambda\rightarrow0$ is also given by $2e\Gamma_{\text{L}}\Gamma_{\text{R}}/(\Gamma_{\text{L}}+\Gamma_{\text{R}})$. For weak electronic-vibrational coupling, vibrational processes (Fig.\ \ref{simpleemissionandabsorption}) do not contribute to the current, 
as they take place on time scales much longer than electronic transport processes, \emph{i.e.} processes that do not include an energy-exchange of the traversing electron with the vibrational mode of the molecular bridge.

\subsubsection{With cooling by electron-hole pair creation processes}
\label{nonanalytic}

For bias voltages in the range $2(\overline{\epsilon}_{1}+\Omega)>e\Phi>2\overline{\epsilon}_{1}$, the vibrational excitation of a mode coupled to a single electronic state remains finite in the limit $\lambda\rightarrow0$. This is illustrated by the gray line in Fig.\ \ref{cureofvibrinst}, which has been obtained for the same parameters as considered above (black line) but with a smaller bias voltage $\Phi=0.45$\,V. Intriguingly, for $\lambda\rightarrow0$, the level of vibrational excitation approaches a non-zero value. 

This result can be derived from the master equation (\ref{genfinalME}). 
For voltages with $2(\overline{\epsilon}_{1}+\Omega)>e\Phi>2\overline{\epsilon}_{1}>\Omega$, we obtain the following set of equations 
\begin{eqnarray}
\label{nonanalyticME}
\Delta^{\nu_{1}} &=&  \Gamma_{\text{L}}  (\nu_{1}+1)\lambda^{2} \rho_{0,0}^{\nu_{1}\nu_{1}}    +  (\nu_{1}+1) \lambda^{2} \Gamma_{\text{L}} \rho^{\nu_{1}+1\nu_{1}+1}_{1,1}   \\
&&+  \Gamma_{\text{R}} (\nu_{1}+1) \lambda^{2} \rho^{\nu_{1}+1\nu_{1}+1}_{1,1} - \Gamma_{\text{R}}  (2\nu_{1}+1) \lambda^{2} \rho^{\nu_{1}\nu_{1}}_{1,1} + \Gamma_{\text{R}}  \nu_{1} \lambda^{2} \rho^{\nu_{1}-1\nu_{1}-1}_{1,1},  \nonumber\\
\Delta^{\nu_{1}} &=&  \Gamma_{\text{L}}  \nu_{1}\lambda^{2} \rho_{1,1}^{\nu_{1}\nu_{1}}  -   \Gamma_{\text{L}} \left( - (2\nu_{1}+1)\lambda^{2} \rho^{\nu_{1}\nu_{1}}_{0,0} + (\nu_{1}+1)\lambda^{2} \rho^{\nu_{1}+1\nu_{1}+1}_{0,0}  \right) ,\nonumber
\end{eqnarray}
with 
\begin{eqnarray}
\Delta^{\nu_{1}}\equiv\Gamma_{\text{L}}\rho_{0,0}^{\nu_{1}\nu_{1}}-\Gamma_{\text{R}}\rho_{1,1}^{\nu_{1}\nu_{1}}.
\end{eqnarray}
From Eqs.\ (\ref{nonanalyticME}), which are valid to second order in $\lambda$, we further deduce
\begin{eqnarray}
0 &=& \left( \nu_{1} \left(2+\Gamma_{\text{L}}/\Gamma_{\text{R}}\right) + \nu_{1} + 1  \right) \rho_{a,a}^{\nu_{1}\nu_{1}}  \\
&&\hspace{-0.5cm} -  (\nu_{1}+1) \lambda^{2} \left( \Gamma_{\text{L}}/\Gamma_{\text{R}} + 2 \right)  \rho^{\nu_{1}+1\nu_{1}+1}_{a,a}  -    \nu_{1} \lambda^{2} \rho^{\nu_{1}-1\nu_{1}-1}_{a,a}, \nonumber
\end{eqnarray}
where $a\in\{0,1\}$. The solution of this equation is given by
\begin{eqnarray}
\rho_{a,a}^{\nu\nu} &=&  \rho_{a,a}^{\nu-1\nu-1} / \left(2+\Gamma_{\text{L}}/\Gamma_{\text{R}}\right), \\
p_{\nu}&=&\rho_{0,0}^{\nu\nu}+\rho_{1,1}^{\nu\nu} \\ 
&=& \frac{\Gamma_{\text{L}}+\Gamma_{\text{R}}}{\Gamma_{\text{L}}+2\Gamma_{\text{R}}} \left(\frac{\Gamma_{\text{R}}}{\Gamma_{\text{L}}+2\Gamma_{\text{R}}}\right)^{\nu}, \nonumber
\end{eqnarray}
which corresponds to an average vibrational excitation of 
\begin{eqnarray}
\langle a^{\dagger}a\rangle= \Gamma_{\text{R}}/\left(\Gamma_{\text{L}}+\Gamma_{\text{R}}\right)\stackrel{\Gamma_{\text{L}}=\Gamma_{\text{R}}}{=}1/2.
\end{eqnarray}
Thus, vibrational excitation 
approaches a finite value as $\lambda\rightarrow0$ and does not vanish, because 
in this case, the leading-order electron-hole pair creation processes are active.

Finally, we note that for even lower bias voltages, $\vert e\Phi\vert<2\overline{\epsilon}_{1}$, the current and the respective current-induced vibrational excitation vanish in the limit $\lambda\rightarrow0$.

\subsubsection{Resonant absorption processes mediated by a higher-lying electronic state}
\label{higherstate}

In this section, we address the question whether resonant absorption processes with respect to a higher-lying electronic state provide a cooling mechanism \cite{Hartle09} that prevents the vibrational instability we observed for a single electronic state (Sec.\ \ref{vibrinstab}). Thereby, we assume the energy $\overline{\epsilon}_{2}$ of the second electronic state to be larger than $\overline{\epsilon}_{1}+\Omega$. 
If the second electronic state would be located within the bias window,
$\vert\overline{\epsilon}_{2}\vert<\vert\overline{\epsilon}_{1}+\Omega\vert<e\Phi$, we would obtain the level of vibrational excitation, which results from transport through this state only, because the other electronic state decouples from the vibrational mode in the limit $\lambda\rightarrow0$.

Model calculations shown by the dashed blue line in Fig.\ \ref{cureofvibrinst} support this conjecture. The current-induced vibrational excitation of a vibrational mode that is coupled to a lower- and a higher-lying electronic state remains finite in the limit $\lambda\rightarrow0$. The parameters of these calculation are the same as those for the black line considered in Sec.\ \ref{vibrinstab}, but include a higher lying-electronic state at $\epsilon_{2}=0.8$\,eV that is coupled to the vibrational mode with a coupling strength of $\lambda_{2}=-0.06$\,eV and to the leads in the same way as the lower-lying state.

This behavior can be rationalized by the master equation, Eq. (\ref{genfinalME}). 
To zeroth order in $\lambda$, we obtain for bias voltages $2\overline{\epsilon}_{2}>e\Phi>2(\overline{\epsilon}_{1}+\Omega)$ 
\begin{eqnarray}
\label{twostatesME}
0 &=& \Gamma_{\text{R},11}  \rho_{10,10}^{\nu_{1}\nu_{1}} -  \rho_{00,00}^{\nu_{1}\nu_{1}} \Gamma_{\text{L},11}  -   \rho_{00,00}^{\nu_{1}\nu_{1}}  \sum_{K\nu_{3}} \Gamma_{K,22} \vert X_{2,\nu_{1}\nu_{3}} \vert^{2} f_{K}(\overline{\epsilon}_{2}+\Omega(\nu_{3}-\nu_{1})),    \\
0 &=& \Gamma_{\text{L},11}   \rho_{00,00}^{\nu_{1}\nu_{1}} -   \rho_{10,10}^{\nu_{1}\nu_{1}} \Gamma_{\text{R},11}   -   \rho_{10,10}^{\nu_{1}\nu_{1}} \sum_{K\nu_{3}} \Gamma_{K,22} \vert X_{2,\nu_{1}\nu_{3}} \vert^{2} f_{K}(\overline{\epsilon}_{2}+\Omega(\nu_{3}-\nu_{1})),    \nonumber\\
0 &=& \sum_{K\nu_{3}}   \Gamma_{K,22}  X^{\dagger}_{2,\nu_{1}\nu_{3}}  X_{2,\nu_{3}\nu_{1}}  \rho_{00,00}^{\nu_{3}\nu_{3}} f_{K}(\overline{\epsilon}_{2}+\Omega(\nu_{1}-\nu_{3})), \nonumber\\
0 &=& \sum_{K\nu_{3}}    \Gamma_{K,22} X^{\dagger}_{2,\nu_{1}\nu_{3}} X_{2,\nu_{3}\nu_{1}}   \rho_{10,10}^{\nu_{3}\nu_{3}} f_{K}(\overline{\epsilon}_{2}+\Omega(\nu_{1}-\nu_{3})), \nonumber
\end{eqnarray}
using the basis functions $\vert 00 \rangle \vert \nu_{1} \rangle$, $\vert 10 \rangle \vert \nu_{1} \rangle$, $\vert 01 \rangle \vert \nu_{1} \rangle$ and $\vert 11 \rangle \vert \nu_{1} \rangle$, respectively.
Thereby, the populations $\rho_{01,01}^{\nu\nu}$ and $\rho_{11,11}^{\nu\nu}$ are treated as second order contributions $\lesssim\lambda^{2}$, because the population of the higher-lying state requires a preceding resonant emission process with respect to the lower-lying state ($\sim\lambda^{2}$). 
Since in the latter two of Eqs.\ (\ref{twostatesME}) all terms are either positive or zero, these equations can only be fulfilled, if the populations $\rho_{00,00}^{\nu\nu}$ and $\rho_{10,10}^{\nu\nu}$ vanish for values of $\nu$ where $f_{K}(\overline{\epsilon}_{2}+\Omega(\nu-\nu_{3}))\neq0$. 
Thus, vibrational levels with a quantum number larger than $(\overline{\epsilon}_{2}-\mu_{\text{L/R}})/\Omega$ are not populated in the limit $\lambda\rightarrow0$. Thus, vibrational excitation is finite in the limit $\lambda\rightarrow0$, 
if a second higher-lying electronic state couples to the vibrational mode with a finite coupling strength $\lambda_{2}$.

\section{Conclusion}

In this article we have studied various limits 
of resonant electron transport through single molecules. 
Thereby, we have focused on limits 
that, due to electronic-vibrational coupling of the molecular bridge, 
exhibit vibrational instabilities, \emph{i.e.}\ an 
infinite level of vibrational excitation.
To this end, we have employed a master equation approach that is based 
on a second order expansion in the molecule-lead coupling. 
To the given order in the molecule-lead coupling, this approach treats 
electronic-vibrational coupling exactly.

An infinite level of vibrational excitation is trivially obtained in the high-temperature limit, $k_{\text{B}}T\rightarrow\infty$, and/or
in the static limit, $\Omega\rightarrow0$, as energy-exchange processes transfer 
the thermal excitation of the leads to 
the vibrational degrees of freedom of the molecular bridge.
A less trivial case is the limit of a large bias voltage, for which we have shown 
that vibrational excitation diverges with increasing bias voltage, $\Phi\rightarrow\infty$.
We have, furthermore, pointed out that this phenomenon results from the suppression of electron-hole pair creation processes. 

Similarly, electron-hole pair creation processes are blocked in the limit of 
vanishing electronic-vibrational coupling, $\lambda\rightarrow0$, 
if the bias voltage is large enough ($e\Phi/2>\overline{\epsilon}_{1}+\Omega$),
and consequently, vibrational excitation diverges in this limit.
For lower bias voltage, where the leading electron-hole pair creation processes are not suppressed, 
we find a finite but non-vanishing level of vibrational excitation in the limit $\lambda\rightarrow0$. 
In the resonant transport regime, 
the vibrational degrees of a molecular junction are thus always excited, 
even if the electronic-vibrational coupling $\lambda$ becomes very weak. 
Cooling mechanisms induced by a higher-lying electronic state, 
as already pointed out in Ref.\ \cite{Hartle09}, or coupling of the vibrational degrees of freedom 
to a thermal bath, as pointed out in Ref.\ \cite{Semmelhack}, however, prevent the vibrational instability 
in the limit $\lambda\rightarrow0$.

Our analysis based on a master equation approach includes no higher order processes, such as co-tunneling processes. In the high-bias regime, 
such processes provide 
an equal number of additional excitation and deexcitation processes. Therefore, 
co-tunneling does not affect 
the corresponding vibrational instability. For a finite bias voltage, however, off-resonant electron-hole pair 
creation processes provide an additional cooling mechanism. The role of these processes 
will be the subject of future research.

\emph{Acknowledgement:}
We thank B.\ Kubala, R.\ Volkovich and U.\ Peskin for helpful 
and inspiring discussions. 
The generous allocation of computing time by the Leibniz 
Rechenzentrum M\"unchen (LRZ) as well as financial support from  
the Deutsche Forschungsgemeinschaft (DFG) and the 
German-Israeli Foundation for Scientific 
Development (GIF) are gratefully acknowledged.
This work was carried
out in the framework of the Cluster of Excellence 'Engineering
of Advanced Materials' of the DFG.


\begin{thebibliography}{78}
\bibitem[{not()}]{noteonMoleculeLeadCoupling}
\bibinfo{note}{Note that the result of Eq.\ (\ref{genfinalME}) is invariant
  with respect to a scaling: $\Gamma_{\text{L/R}}\rightarrow \alpha
  \Gamma_{\text{L/R}}$, where $\alpha\in\mathbb{R}$.}
\end{thebibliography}
\end{document}